\documentclass [12pt]{article}
\usepackage[dvips]{graphicx}

\usepackage{epsfig}
\usepackage{graphicx}
\usepackage{indentfirst}
\usepackage{amsmath}
\usepackage{amsfonts}
\usepackage{amssymb}
\usepackage{hyperref}
 \usepackage{latexsym}
\usepackage{color}

\begin{document}
\begin{center}
{\Large\bf  Late time cosmological approach in mimetic $f(R,T)$ gravity }\\

\medskip

  E. H. Baffou$^{(a)}$\footnote{e-mail:baffouhet@gmail, baffouh.etienne@yahoo.fr},  
 M. J. S. Houndjo$^{(a,b)}$\footnote{e-mail:
 sthoundjo@yahoo.fr}, M. Hamani-Daouda$^{(c)}$\footnote{e-mail:daoudah8@yahoo.fr} and F. G. Alvarenga$^{(d)}$\footnote{e-mail:f.g.alvarenga@gmail.com }
 $^a$ \,{\it Institut de Math\'{e}matiques et de Sciences Physiques (IMSP)}\\
   {\it 01 BP 613,  Porto-Novo, B\'{e}nin}\\
  $^{b}$\,{\it Facult\'e des Sciences et Techniques de Natitingou - B\'enin} \\
	$^{c}$\,{\it D\'epartement de Physique - Universit\'e de Niamey, Niamey, Niger}\\
	$^{d}$\,{\it Departamento de Engenharia e Ci\^encias Naturais - CEUNES - Universidade\\
	Federal do Esp\'irito Santo - CEP $29933-415$ - S\~ao Mateus/ ES, Brazil}
\end{center}
\begin{abstract}
In this paper, we investigate the late-time cosmic acceleration in mimetic  $f(R,T)$ gravity with Lagrange multiplier and potential
in a Universe containing, besides radiation  and dark energy, a
self-interacting (collisional) matter. We obtain through the modified Friedmann equations, the main equation that can describe the cosmological evolution and with several models from $\mathcal{Q}(z)$ and  the well known particular model $f(R,T)$,
we perform an analysis of the late-time evolution. We examine the behavior of the  Hubble parameter, the dark energy equation of state and the total effective equation of state and we  compare in each case the resulting picture with the non-collisional matter (assumed as dust)
and also with the collisional matter in mimetic $f(R,T)$ gravity. The results obtained  are in good agreement with the observational
data and show that in presence of the collisional matter the dark energy oscillations in mimetic $f(R,T)$ gravity can be damped.
\end{abstract}

 Pacs numbers: 04.50.Kd, 95.36.+x, 98.80.-k \\
%
 \vspace{0.2cm}

\section{Introduction}
One of the most  important recent scientific discoveries is provided by a set of observational data, indicated on late time acceleration
expansion of the whole Universe as well as the initial era, the inflationary epoch \cite{baffoua}.
It is noteworthy that the accelerated expansion of the Universe is created from a mysterious energy called dark energy,
which is dominated component. Modified gravity is one of the methods to describe the accelerated expansion of the Universe.
Modified gravity can be considered a new challenge to cure shortcomings of General Relativity at infrared and ultraviolet scales.
They are an approach that, by preserving the undoubtedly positive results of Einstein’s Theory, is aimed to address conceptual and experimental
problems recently emerged in Astrophysics, Cosmology and High Energy Physics. 
In particular, the goal is to encompass, in a self-consistent scheme, problems like Inflation, dark energy, dark
matter, large scale structure and, first of all, to give at least an effective description of quantum gravity.
The late-time cosmic acceleration can, in principle from a modification of gravity rather than an exotic source of matter with a negative pressure.
A lot of works on modified gravity have be done to identify the origin of dark energy \cite{baffou5} over the last years.
The attractive point in modified gravity models is that they are generally more strongly constrained from cosmological observations and  local gravity
experiments than the models based on the exotic source of matter.
One of the simplest modifications to GR  is the $f(R)$ theories of gravity in which the Lagrangian density is supposed to an arbitrary function $R$
\cite{baffou6}. For informative reviews and very important papers on these theories, see \cite{baffou8}-\cite{baffou16}.
In recent years, a new theory was developed in Ref.\cite{baffou17}, named
$f(R,T)$ gravity, that can also be considered as a generalization of $f(R)$ gravity. In this theory, an arbitrary function
of the Ricci scalar $R$ and the trace of the energy-momentum tensor is introduced instead of an arbitrary function of only
the Ricci scalar. The main justifications for employing the trace of the energy-momentum tensor may be induced by exotic imperfect
fluids or quantum effects (conformal anomaly).
Different aspects of such theory have been investigated in the literature \cite{baffou18}-\cite{baffou25}.
Recently, another kind of alternative gravity model has been proposed, for
which the metric is not considered a fundamental quantity. Instead, it is taken
as a function of an auxiliary metric $ \tilde{g}_{\mu\nu}$ and a scalar field $\phi$, contemplating the so called Mimetic Gravity (MG) \cite{baffou26}.
Note that such a dependence of the metric makes the application of the variational principle in the model action to yield
more general equations of motion (EoM) than that of purely Einsteinian relativity theory.
To unify $f(R)$ gravity with this very interesting mimetic theory,
it has been proposed mimetic $f(R)$ gravity \cite{baffou27} as a new class of modified gravities
with the same inspiration as mimetic theory.
Very recently the author \cite{baffou28} demonstrated how the $f(R)$ gravity, the mimetic potential and the Lagrange multiplier affect
the late-time cosmological evolution and in Ref.\cite{baffou29}, it was demonstrated that in the context of mimetic $f(R)$ gravity with Lagrange multiplier and mimetic
potential, it is possible to solve in a kind of elegant way the problem of dark energy oscillations at late times. For several authors works doing in mimetic
modified gravity, see \cite{baffou30}-\cite{baffou35}.
In this paper, we adopt the mimetic $f(R,T)$  gravity
approach with scalar potential $V(\phi)$ and with Lagrange multiplier $\lambda(\phi)$  to describe the late-time cosmological evolution and dark energy eras.
The specific main in this paper is to extend the works done by the authors \cite{baffou36} in mimetic $f(R,T)$ gravity context. 
We shall investigated how the $f(R,T)$ gravity in presence of the potential and Lagrange
multiplier can  offers much freedom for realizing various cosmic evolution scenarios and can be allowed to have compatibility with observational data.
The present paper is organized as follows: In Sec {\ref{sec1}}, we briefly review the  mimetic cosmology in $f(R,T)$ gravity.
The Sec {\ref{sec2}} is devoted to the study of the late-time cosmological evolution in mimetic $f(R,T)$ gravity.
Our conclusion are presented in the last section.

\section{ Brief Review in mimetic $f(R,T)$ Gravity Models } {\label{sec1}}
We provide in this section a brief review of mimetic $f(R,T)$ gravity with Lagrange multiplier and potential. 
The main idea to study the mimetic approach in modified gravity comes from the general class of mimetic gravities \cite{baffou26},\cite{a2}-\cite{a8} 
is that parametrizing the metric using new degrees of freedom one can obtain
modified field equations which may admit a wider family of solutions. Therefore, we express the physical metric $g_{\mu\nu}$ in
terms of an auxiliary metric $\tilde{g}_{\mu\nu}$ and of an auxiliary scalar field $\phi$, as follows, 
\begin{eqnarray}
g_{\mu\nu}= -\tilde{g}^{\rho\sigma}\partial_{\rho}\phi \partial_{\sigma}\phi \tilde{g}_{\mu\nu} \label{1}, 
\end{eqnarray}
and thus the gravitational field variation will be performed in terms of both the auxiliary metric \^{g}$_{\mu\nu}$ and auxiliary scalar field $\phi$.
Eq.(\ref{1}) shows
\begin{eqnarray}
g^{\mu\nu}(\tilde{g}_{\mu\nu},\phi) \partial_{\mu}\phi\partial_{\nu}\phi = -1. \label{2}
\end{eqnarray}
Since we are interested to investigate the cosmological implications of mimetic $f(R,T)$
gravity,  we consider in the following  the flat Friedmann- Robertson-Walker (FRW) metric which the line element
\begin{eqnarray}
 ds^{2}= dt^{2}-a(t)^{2}[dx^{2}+ dy^{2}+dz^{2}],\label{3}
\end{eqnarray}
where $a(t)$ is the scale factor.
We assume in mimetic $f(R,T)$ gravity with Lagrange multiplier $\lambda(\phi)$ and mimetic potential $V(\phi)$, the gravitionnal action coupled with matter
as
\begin{eqnarray}
S =  \int \sqrt{-g} dx^{4} \Big[f \big(R(g_{\mu\nu}),T \big)-V(\phi) + \lambda(g^{\mu\nu}\partial_{\mu}\phi \partial_{\nu}\phi+1) +\mathcal{L}_m \Big],
\label{4}
\end{eqnarray}
where we assumed $16\pi G = 1$; $R= g^{\mu\nu}R_{\mu\nu}$ and $T= g^{\mu\nu}T_{\mu\nu}$ denotes respectively the curvature scalar of Ricci tensor $R_{\mu\nu}$ and the trace of energy-momentum tensor 
$T_{\mu\nu}$; $\mathcal{L}_m$ being the matter Lagrangian density of all fluids present. \\
We defined the energy-momentum tensor of the matter from the  matter Lagrangian density $\mathcal{L}_m$ as
\begin{eqnarray}
T_{\mu\nu}=-\frac{2}{\sqrt{-g}}\frac{\delta\left(\sqrt{-g}\mathcal{L}_m\right)}{\delta g^{\mu\nu}}.\label{5}
\end{eqnarray} 
Variation of the action (\ref{4}) with respect to the tensor metric $g_{\mu\nu}$ is given by
\begin{eqnarray}
\frac{1}{2} g_{\mu\nu}f(R,T) -R_{\mu\nu}f_{R}+\nabla_\mu \nabla_{\nu}f_R-g_{\mu\nu}\Box f_{R}+\frac{1}{2}g_{\mu\nu}\bigg(-V(\phi)+ 
\lambda(g^{\rho\sigma}\partial_{\rho}\phi \partial_{\sigma}\phi+1)\bigg)- \nonumber \\ \lambda \partial_{\mu}\phi \partial_{\nu}\phi - f_{T}(T_{\mu\nu}+\Theta_{\mu\nu})
+\frac{1}{2} T_{\mu\nu}=0,\label{6}
\end{eqnarray}
where
\begin{eqnarray}
\Theta_{\mu\nu}\equiv g^{\alpha\beta}\frac{\delta T_{\alpha \beta}}{\delta g^{\mu\nu}}=-2T_{\mu\nu}+g_{\mu\nu}\mathcal{L}_m
-2g^{\alpha\beta}\frac{\partial^2 \mathcal{L}_m}{\partial g^{\mu\nu}\partial g^{\alpha \beta}}\label{7},
\end{eqnarray}
$\nabla_i$ denotes the covariant derivative with respect to the metric $g_{\mu\nu}$ and $\Box = \nabla^{i}\nabla_i$, the d'Alembertian operator. 
In Eqs.(\ref{6}), $f_{R}$ and $f_{T}$ represents the partial derivation of $ f(R,T) $ with respect to the $ R $, $T$, respectively.
Notice that the auxiliary metric $\tilde{g}_{\mu\nu}$ does not appear in these equations  by itself but
only via the physical metric $g_{\mu\nu}$ , while the scalar field $\phi$ enters the equations
explicitly.\\
At follows, we assume that the  matter content of the universe as the perfect fluid for which the energy-momentum tensor is expressed as
\begin{eqnarray}
T_{\mu\nu} = (\rho_{matt}+ p_{matt}) u_{\mu}u_{\nu}+ p_{matt} g_{\mu\nu} \label{8}, 
\end{eqnarray}
where $\rho_{matt}$ and $p_{matt}$ are the energy density and the pressure of the
matter, respectively, $u_{\mu}$ is the four-velocity. In this way, the  matter Lagrangian density can be chosen as $\mathcal{L}_m = - p_{matt}$.
In fact, the field equations (\ref{6}) yields
\begin{eqnarray}
 \frac{1}{2} g_{\mu\nu}f(R,T) -R_{\mu\nu}f_{R}+\nabla_\mu \nabla_{\nu}f_R-g_{\mu\nu}\Box f_{R}+\frac{1}{2}g_{\mu\nu}\bigg(-V(\phi)+ 
\lambda(g^{\rho\sigma}\partial_{\rho_m}\phi \partial_{\sigma}\phi+1)\bigg)- \nonumber \\ \lambda \partial_{\mu}\phi \partial_{\nu}\phi + 
f_{T}(T_{\mu\nu}+ p_mg_{\mu\nu})
+\frac{1}{2} T_{\mu\nu}=0. \label{9}
\end{eqnarray}
Varying the gravitational action (\ref{4}) with respect to the auxiliary scalar field $\phi$, one gets
\begin{eqnarray}
-2\nabla^{\mu}(\lambda\partial_{\mu}\phi)-  V'(\phi)= 0 \label{10},
\end{eqnarray}
where the prime denote the derivative of the mimetic potential with respect to the auxiliary scalar $\phi$.
On the other hand, by the variation with respect to the Lagrange multiplier $\lambda$, we obtain
\begin{eqnarray}
g^{\mu\nu}\partial_{\mu}\phi\partial_{\nu}\phi = -1,\label{11} 
\end{eqnarray}
which shows that the scalar field will not be a propagating degree of freedom \cite{baffou30}.
This equation express the constraint equation (\ref{2}) in mimetic gravity, equation obtained by varying the gravitational action  
with respect to the Lagrange multiplier $\lambda$.\\
Considering the FRW space-time (\ref{3}) and assuming that $\phi$ depends only on time coordinate $t$, the field equations (\ref{9}), (\ref{10}) and (\ref{11})
are written as
\begin{eqnarray}
-f(R,T) +6(\dot{H}+H^2)f_{R}-6H\frac{df_{R}}{dt}-\lambda(\dot{\phi}^2+1)+V(\phi)+\rho_{matt}(2f_{T}+1)-2f_{T}p_{matt} = 0, \label{12} 
\end{eqnarray}
\begin{eqnarray}
f(R,T) -2(\dot{H}+3H^2)f_{R}+ 4H\frac{df_{R}}{dt}+2\frac{d^2f_{R}}{dt^2} -\lambda(\dot{\phi}^2-1)-V(\phi)+(4f_{T}+1)p_{matt} = 0, \label{12a} 
\end{eqnarray}
\begin{eqnarray}
2\frac{d(\lambda \dot{\phi})}{dt}+6H\lambda \dot{\phi}-V'(\phi) = 0, \label{13}  
\end{eqnarray}
\begin{eqnarray}
\dot{\phi}^2 -1 = 0. \label{14} 
\end{eqnarray}
In these expressions,  the ``dot'' represents the derivative
 with respect to the cosmic time $t$ whereas the ``prime'' denotes the derivative with respect to the
auxiliary scalar $\phi$. From  the eq. (\ref{14}), we remark that $\phi$ can be identified as the time coordinate $(\phi = t)$. Then
Eq.(\ref{12}) reduces to
\begin{eqnarray}
f(R,T)-2(\dot{H}+3H^2)f_{R}+ 4H\frac{df_{R}}{dt}+2\frac{d^2f_{R}}{dt^2} - V(t) +(4f_{T}+1)p_{matt} = 0. \label{15}  
\end{eqnarray}
In $f(R,T)$ gravity, the mimetic potential $V(t)$ can be expressed as
\begin{eqnarray}
V(t) = 2\frac{d^2f_{R}}{dt^2}+4H\frac{df_{R}}{dt}+f(R,T)-2(\dot{H}+3H^2)f_{R}+(4f_{T}+1)p_{matt}. \label{16}   
\end{eqnarray}
Within specific forms of the $f(R,T)$ models and the Hubble parameter, the correspondent mimetic potential can be found. Once this expression known,
Eq. (\ref{12}) can be solved with respect to Lagrange multiplier $\lambda (t)$
\begin{eqnarray}
\lambda(t) = -\frac{1}{2}f(R,T)+ 3(\dot{H}+H^2)f_{R}-3H\frac{df_{R}}{dt}+\rho_{matt} (f_{T}+\frac{1}{2})-f_{T}p_{matt}+\frac{1}{2} V(t).  \label{17}
\end{eqnarray}
\section{Late time cosmological evolution in mimetic $f(R,T)$ gravity}{\label{sec2}}

In order to establish the main differential equation which governs the dark energy oscillations evolution, we recast the FRW equation (\ref{12}) as follows,
\begin{eqnarray}
3H^2f_{R} = \rho_{matt}(f_{T}+\frac{1}{2})-f_{T}p_{matt}+\frac{1}{2}(Rf_{R}-f(R,T)) +\frac{V(t)-2\lambda(t)}{2}        -3H\dot{f_{R}} = 0 \label{18}, 
\end{eqnarray}
where $\rho_{matt}$ and $p_{matt}$ are the total energy-density and pressure of all fluids present in Universe. 
In $f(R,T)$ gravity,  the trace $T$ of the energy-momentum tensor depends on the nature of the matter content. As novelty, 
we assume in this paper, the matter content of the universe as collisional matter and relativistic matter (radiation).  
The model of collisional matter has been studied in some works,
leading to interesting results \cite{baffou36},\cite{b2}-\cite{b4}. This approach of considering forms of matter other than cold dark matter can
educate us on the choice of the models of modified gravity.
Accordingly, we write the total energy density for the mimetic  $f(R,T)$ gravity case as
\begin{eqnarray}
\rho_{matt}= \varepsilon_m + \rho_{r0} a^{-4}  \label{19},
\end{eqnarray}
where $\varepsilon_m$ is the energy density of collisional matter given by
\begin{eqnarray}
\varepsilon_{m} = \rho_{m0}a^{-3} \Bigg (1+ \Pi_0 + 3w \ln(a)\Bigg) \label{20},
\end{eqnarray}
and $\rho_{r0}$ the current energy density of radiation, $\rho_{m0}$ and $\Pi_0$ denote present values of the motion invariant mass energy density and of
the potential energy, respectively.
We can reformulate the total matter energy density (\ref{19}) in term of the parameter $\mathcal{G}(a)$ describing the nature of the collisional
matter (view as perfect fluid) as
\begin{eqnarray}
\rho_ {matt} = \rho_{m0} \bigg( \mathcal{G}(a)+ \chi a^ {-4}\bigg), \label{20a}
\end{eqnarray}
where the parameter $\mathcal{G}(a)$ is equal  to
\begin{eqnarray}
\mathcal{G}(a) = a^{-3} \bigg(1+\Pi_0+3w \ln(a) \bigg). \label{20b}      
\end{eqnarray}
Note that for  the non-collisional matter (assumed as the dust), for which the parameter $w=0$, one gets  $ \mathcal{G}(a)= a^{-3} $.
Eq.(\ref{18}) can be reformulated as
\begin{eqnarray}
H^2 +(1-f_{R})(H\frac{dH}{d\ln a} +H^2 )+ \frac{1}{6}\bigg(f(R,T)-R\bigg)+ H^2f_{RR}\frac{dR}{d\ln a}-\nonumber \\ \frac{1}{3}f_{T}(\rho_{matt}-p_{matt}) -
\frac{V(t)-2\lambda(t)}{6} =\frac{\rho_{matt}}{6}, \label{21} 
\end{eqnarray}
while the scalar curvature $R$ can be expressed as
\begin{eqnarray}
R = 12 H^2 +6H\frac{dH}{d\ln a}. \label{22}
\end{eqnarray}
We introduce for the reason of the simplicity the following function of the redshift $z$ ,
\begin{eqnarray}
 \mathcal{Q}(a(z)) = V(a(z)) -2\lambda (a(z)) \label{23}  ,
\end{eqnarray}
which depends on the mimetic potential and on the Lagrange multiplier.
In terms of the parameters $\mathcal{G}(a)$ and $Q(a)$, Eq.(\ref{21}) can be rewritten as
\begin{eqnarray}
H^2 +(1-f_{R})(H\frac{dH}{d\ln a} +H^2 )+ \frac{1}{6}\bigg(f(R,T)-R\bigg)+ H^2f_{RR}\frac{dR}{d\ln a} -\nonumber \\
\rho_{m0}\bigg(\mathcal{G}(a)+ \chi a^ {-4}\bigg)\bigg(\frac{1}{6}+\frac{1}{3}f_{T}(1-w)\bigg)-\frac{\mathcal{Q}(a)}{6}= 0. \label{23a}
\end{eqnarray}

In an effort to better study the late time cosmological evolution in mimetic $f(R,T)$ gravity, we may introduce the following variable
\begin{eqnarray}
y_H \equiv \frac{\rho_{DE}}{\rho_{m0}}= \frac{H^2}{ {\bar m}^{2}}-\mathcal{G}(a)-\mathcal{Q}(a)- \chi a^{-4} , \label{24}
\end{eqnarray}
\begin{eqnarray}
y_R \equiv  \frac{R}{ {\bar m}^{2} }- \frac{d\mathcal{G}(a)}{d\ln a}-\frac{d\mathcal{Q}(a)}{d\ln a},\label{25} 
\end{eqnarray}
where $\rho_{DE}$ denotes  the energy density of dark energy, ${\bar m}^{2}$ being the mass scale
and $\chi$ the ratio defined as $\chi = {\rho_{r0}}/{\rho_{m0}}$. The dark energy scale $y_H$ is the new variable that can be described the late time 
cosmological evolution.
Making use of (\ref{23a}), the expression $\frac{1}{{\bar m}^{2}} \frac{dR} {d\ln a}$ yields
\begin{eqnarray}
\frac{1}{{\bar m}^{2}} \frac{dR} {d\ln a}= \frac{1}{H^2 f_{RR}} \Bigg [ \bigg(\mathcal{G}(a)+\chi a^{-4}\bigg)\Bigg(\frac{\rho_{m0}}{{\bar m}^{2}}\bigg(\frac{1}{6}+\frac{1}{3}f_T(1-w)\bigg)-1\Bigg)\nonumber \\
-\frac{1}{6{\bar m}^{2}}\bigg(f(R,T)-R\bigg)-y_H +\mathcal{Q}(a)\bigg(\frac{1}{6{\bar m}^{2}}-1\bigg)-(1-f_R)\bigg(\frac{H}{{\bar m}^{2}}\frac{dH}{d\ln a}+\frac{H^2}{{\bar m}^{2}} \bigg) \Bigg].
\label{26}
\end{eqnarray}
Combining the differentiation of Eq.(\ref{25}) with respect to $\ln a$  with Eq. (\ref{26}), we obtain
\begin{eqnarray}
&&\frac{dy_R}{d\ln a}= -\frac{d^2 \mathcal{G}(a)}{d\ln a^2}-\frac{d^2 \mathcal{Q}(a)}{d\ln a^2}+\nonumber\\
&&\frac{1}{{\bar m}^{2}\bigg(y_H+\mathcal{G}(a)+{\mathcal{Q}a(z)}+\chi a^{-4}\bigg) f_{RR}} \Bigg [ \bigg(\mathcal{G}(a)+\chi a^{-4}\bigg)\Bigg(\frac{\rho_{m0}}{{\bar m}^{2}}\bigg(\frac{1}{6}+\frac{1}{3}f_T(1-w)\bigg)-1\Bigg)
\nonumber \\
&&-\frac{1}{6{\bar m}^{2}}\bigg(f(R,T)-R\bigg) +\mathcal{Q}(a)\bigg(\frac{1}{6{\bar m}^{2}}-1\bigg)-y_H - \nonumber \\
&&(1-f_R)\bigg(\frac{1}{2}\frac{dy_H}{d\ln a}+\frac{1}{2}\frac{d\mathcal{G}(a)}{d\ln a}+\frac{1}{2}\frac{d\mathcal{Q}(a)}{d\ln a} +y_H +\mathcal{G}(a)+\mathcal{Q}(a)-\chi a^{-4} \bigg) \Bigg].
\label{27}
\end{eqnarray}
Moreover, the curvature scalar (\ref{22}) can be expressed as
\begin{eqnarray}
R= 3 {\bar m}^{2}\Bigg[4y_{H}+4\mathcal{G}(a)+4\mathcal{Q}(a)+\frac{dy_H}{d\ln a}+\frac{d\mathcal{G}(a)}{d\ln a}+\frac{d\mathcal{Q}(a)}{d\ln a} \Bigg]. \label{28}
\end{eqnarray}
Upon differentiation of Eq. (\ref{24}) with respect to $\ln a$ we obtain
\begin{eqnarray}
 \frac{dy_H}{d\ln a} = \frac{2H}{ {\bar m}^{2}}\frac{dH}{d \ln a} -\frac{d\mathcal{G}(a)}{d\ln a}-\frac{d\mathcal{Q}(a)}{d\ln a}+4\chi a^{-4}. \label{29}
\end{eqnarray}
Using Eqs.(\ref{22}), (\ref{24}) and (\ref{25}), Eq.(\ref{29}) becomes
\begin{eqnarray}
\frac{dy_H}{d\ln a} = \frac{1}{3} y_R- 4 y_H -\frac{2}{3}\frac{d \mathcal {G}(a)}{d\ln a}-\frac{2}{3}\frac{d\mathcal{Q}(a)}{d\ln a}-4\mathcal{G}(a)-4\mathcal{Q}(a). \label{30}
\end{eqnarray}
By operating the differentiation of the relation (\ref{30}) with respect to $\ln a $ and also by using the Eq.(\ref{27}), we obtain the following differential equation
\begin{eqnarray}
&&\frac{d^2 y_H}{d\ln a ^2}+ \bigg(4+\frac{1-f_R}{6{\bar m}^{2}f_{RR}(y_H+\mathcal{G}(a)+\mathcal{Q}(a)+\chi a^{-4})}\bigg) \frac{dy_H}{d\ln a}+\nonumber \\
&&\bigg(\frac{2-f_R}{3{\bar m}^{2}f_{RR}(y_H+\mathcal{G}(a)+\mathcal{Q}(a)+\chi a^{-4})}\bigg)y_H+ \mathcal{P}(a)=0 \label{31}, 
\end{eqnarray}
where $\mathcal{P}(a)$ is given by
\begin{eqnarray}
&&\mathcal{P}(a) = \frac{d^2 \mathcal{G}(a)}{d\ln a ^2} +\frac{d^2 \mathcal{Q}(a)}{d\ln a ^2}+4\frac{d \mathcal{G}(a)}{d\ln a}+4\frac{d \mathcal{Q}(a)}{d\ln a}+\nonumber \\
&&\frac{1}{18{\bar m}^{2}\bigg(y_H+\mathcal{G}(a)+{\mathcal{Q}(a)}+\chi a^{-4}\bigg) f_{RR}} \Bigg [ \bigg(\mathcal{G}(a)+\chi a^{-4}\bigg)\Bigg(\frac{\rho_{m0}}{{\bar m}^{2}}\bigg(-1+2f_T(w-1)\bigg)\Bigg)
\nonumber \\
&&+\frac{1}{{\bar m}^{2}}\bigg(f(R,T)-R\bigg) +\bigg(\mathcal{Q}(a)+\mathcal{G}(a)\bigg)(12-6f_R)- \frac{1}{{\bar m}^{2}}\mathcal{Q}(a) +6\chi a^{-4}f_R - \nonumber \\
&&3(f_R-1)\bigg(\frac{d\mathcal{G}(a)}{d\ln a}+\frac{d\mathcal{Q}(a)}{d\ln a} \bigg) \Bigg]. \label{32}
\end{eqnarray}
Taking into account the relations
\begin{eqnarray}
\frac{d}{d\ln a}=-(1+z)\frac{d}{dz}, \label{58} 
\end{eqnarray}
\begin{eqnarray}
\frac{d^2}{{d\ln a}^{2}} = {(1+z)}^{2}\frac{d^2}{{dz}^{2}}+(1+z)\frac{d}{dz},  \label{59} 
\end{eqnarray}
we can easily express  all the physical quantities of (\ref{32}) in terms of the redshift $z$, as follows
\begin{eqnarray}
\frac{d^2 y_H}{dz^2}+\frac{1}{1+z}\bigg(-3+\frac{f_R-1}{6{\bar m}^{2}f_{RR}(y_H+\mathcal{G}(z)+\mathcal{Q}(z)+\chi{(1+z)}^{4})}\bigg) \frac{dy_H}{dz} 
+\nonumber \\
\frac{1}{{(1+z)}^2}\bigg[\frac{2-f_R}{3{\bar m}^{2}f_{RR}(y_H+\mathcal{G}(z)+\mathcal{Q}(z)+\chi {(1+z)}^{4})}\bigg]y_H+ \mathcal{P}(z)=0, \label{31}
\end{eqnarray}
where
\begin{eqnarray}
&& \mathcal{P}(z)= \frac{d^2\mathcal{G}(z)}{dz^2}+ \frac{d^2\mathcal{Q}(z)}{dz^2}-\frac{3}{1+z}\bigg(\frac{d\mathcal{G}(z)}{dz}+\frac{d\mathcal{Q}(z)}{dz}\bigg)
+\nonumber \\
&&\frac{1}{18{(1+z)}^2{\bar m}^{2}\bigg(y_H+\mathcal{G}(z)+{\mathcal{Q}(z)}+\chi{(1+z)}^{4}\bigg) f_{RR}} \times \nonumber \\
&&\Bigg[ \bigg(\mathcal{G}(z)+\chi{(1+z)}^{4}\bigg)\Bigg(\frac{\rho_{m0}}{{\bar m}^{2}}\bigg(-1+2f_T(w-1)\bigg)\Bigg)
\nonumber \\
&&+\frac{1}{{\bar m}^{2}}\bigg(f(R,T)-R\bigg) +\bigg(\mathcal{Q}(z)+\mathcal{G}(z)\bigg)(12-6f_R)- \frac{1}{{\bar m}^{2}}\mathcal{Q}(z) +6\chi {(1+z)}^{4}f_R 
+\nonumber \\
&&3(1+z)(f_R-1)\bigg(\frac{d\mathcal{G}(z)}{dz}+\frac{d\mathcal{Q}(z)}{dz} \bigg) \Bigg], \label{32}
\end{eqnarray}
\begin{eqnarray}
g(z)= {(1+z)}^3\bigg(1+\Pi_0-3w\ln(1+z)\bigg). \label{33} 
\end{eqnarray}
Inserting Eq.(\ref{33}) in differential equation (\ref{31}), one have
\begin{eqnarray}
\frac{d^2 y_H}{dz^2}+\frac{1}{1+z}\Bigg[-3+\frac{f_R-1}{6{\bar m}^{2}f_{RR}\bigg(y_H+{(1+z)}^3\big(1+\Pi_0-3w\ln(1+z)\big)+\mathcal{Q}(z)+\chi{(1+z)}^{4}\bigg)}\Bigg] \frac{dy_H}{dz} 
\nonumber \\
+\frac{1}{{(1+z)}^2}\Bigg[\frac{2-f_R}{3{\bar m}^{2}f_{RR}\bigg(y_H+ {(1+z)}^3\big(1+\Pi_0-3w\ln(1+z)\big)+\mathcal{Q}(z)+\chi {(1+z)}^{4}}\Bigg]y_H+ \mathcal{P}(z)=0,  
\label{34} 
\end{eqnarray}
and the corresponding function $\mathcal{P}(z)$ reads

\begin{eqnarray}
&& \mathcal{P}(z)=  \frac{d^2\mathcal{Q}(z)}{dz^2}-\frac{3}{1+z}\frac{d\mathcal{Q}(z)}{dz}-3(1+z)(1+\Pi_0+2w)
+\nonumber \\
&&\frac{1}{18{(1+z)}^2{\bar m}^{2}\bigg(y_H+ {(1+z)}^3\big(1+\Pi_0-3w\ln(1+z)\big)   +{\mathcal{Q}(z)}+\chi{(1+z)}^{4}\bigg) f_{RR}} \times \nonumber \\
&&\Bigg[ \bigg({(1+z)}^3\big(1+\Pi_0-3w\ln(1+z)\big)  +\chi{(1+z)}^{4}\bigg)\Bigg(\frac{\rho_{m0}}{{\bar m}^{2}}\bigg(-1+2f_T(w-1)\bigg)\Bigg)
\nonumber \\
&&+\frac{1}{{\bar m}^{2}}\bigg(f(R,T)-R\bigg) +\bigg(\mathcal{Q}(z)+{(1+z)}^3\big(1+\Pi_0-3w\ln(1+z)\big)\bigg)(12-6f_R) \nonumber\\
&&- \frac{1}{{\bar m}^{2}}\mathcal{Q}(z) +6\chi {(1+z)}^{4}f_R +3(1+z)(f_R-1)\frac{d\mathcal{Q}(z)}{dz} \nonumber \\
&& +9{(1+z)}^3(f_R-1)(1+\Pi_0-w-3w\ln(1+z))           \Bigg]. \label{35}
\end{eqnarray}

By looking the main equation (\ref{34}), it is obvious that it describe the late time cosmological evolution of the dark energy in a Universe 
filled with collisional matter and radiation and is strongly affected by mimetic potential and the Lagrange multiplier which are contained in 
the function $\mathcal{Q}(z)$. At follows, we shall specify the exact form of the function $Q(z)$ and the $f(R,T)$ models to solve numerically the differential
equation (\ref{34}). Once this equation be solved, we will perform the late time  evolution of the cosmological parameters in mimetic $f(R,T)$  gravity.

\section{Numerical Analysis for variable $\mathcal{Q}(z)$ models} {\label{sec3}}
In this section, we perform the numerically analysis of Eq.(\ref{34}) to study the late time cosmological evolution in mimetic $f(R,T)$ gravity.
To do,  we focus our attention in particular viable  $f(R,T) = f(R)+f(T)$ models  where $f(R) = R-2\Lambda (1-e^{\frac{R}{b\Lambda}})$ \cite{nojori} 
and $f(T) =T^{\beta}$ \cite{baffou36}. In present $f(R,T)$ models considered, $\Lambda$ represents the present time cosmological constant, $b$ is a positive free parameter
which is assumed to be $\mathcal{O}(1)$ and $\beta$ a real constant. With variable models of $\mathcal{Q}(z)$, we plot in mimetic $f(R,T)$ gravity, the
evolution of the Hubble parameter $H(z)$, the total effective equation of state $w_{eff}$, 
the  parameter of equation of state for dark energy $w_{DE}$  versus $z$ and we compare the results obtained in mimetic $f(R,T)$ gravity in presence
of the collisional matter $ ( w = 0.6)$ with those in presence of the non-collisional matter $(w =0 )$.
Respectively, these parameters can be expressed as follows

\begin{eqnarray}
H(z) = \sqrt{{\bar m}^{2}\bigg(y_H+\mathcal{Q}(z)+{(1+z)}^3\big(1+\Pi_0-3w\ln(1+z)\big)+\chi{(1+z)}^{4}\bigg)} \label{66} 
\end{eqnarray}
\begin{eqnarray}
w_{eff}(z) = -1+\frac{2(1+z)}{3H(z)} \frac{dH(z)}{dz}.\label{68} 
\end{eqnarray}
\begin{eqnarray}
w_{DE}(z) = -1+\frac{1}{3}(1+z)\frac{1}{y_H}\frac{d y_H}{dz},\label{64}
\end{eqnarray}

\begin{figure}[h]
  \centering
  \begin{tabular}{rl}
  \includegraphics[width=7cm, height=7cm]{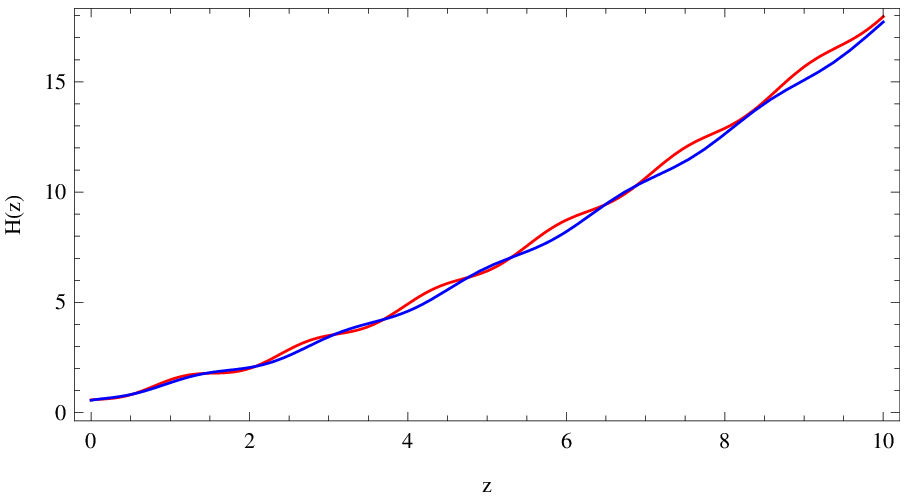}&
\includegraphics[width=7cm, height=7cm]{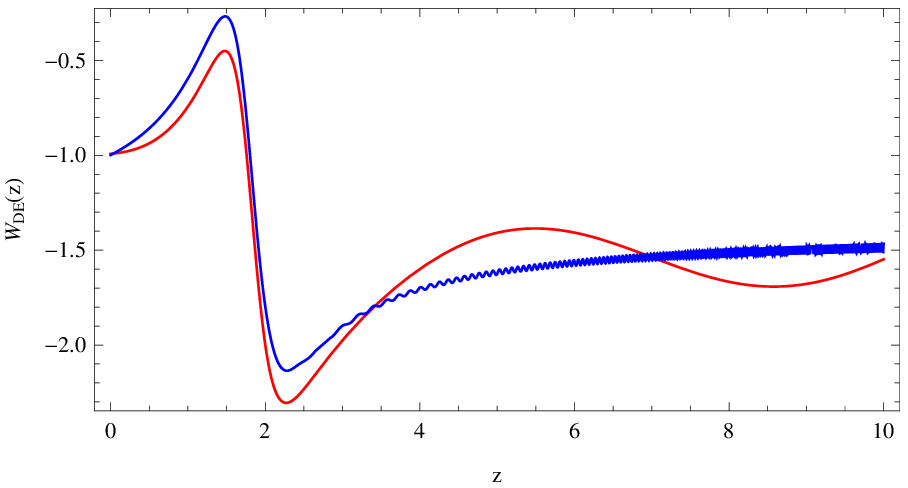}
\end{tabular}
\label{fig1}
\end{figure}
  
\begin{figure}[h]
\centering
\begin{tabular}{rl}
\includegraphics[width=7cm, height=7cm]{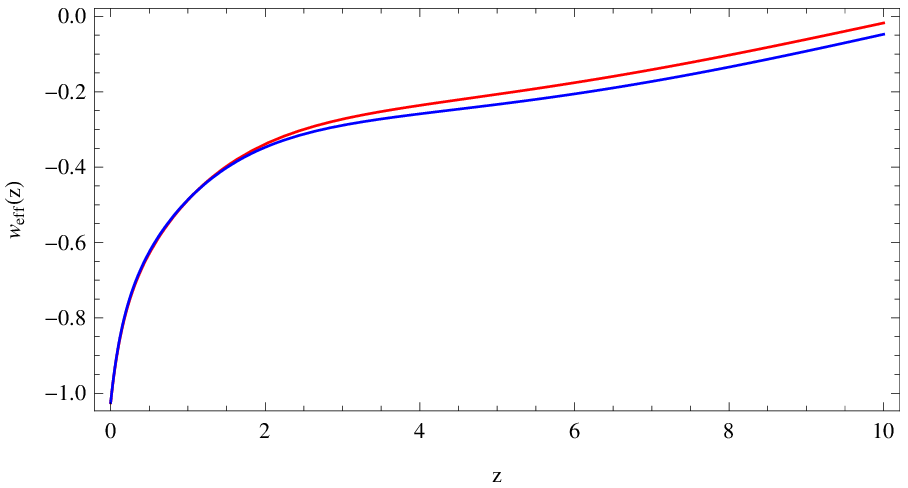}
\end{tabular}
\caption{Comparison of the Hubble parameter $H(z)$ versus $z$, of the dark energy equation of state parameter $w_{DE}$ versus $z$ and 
of the effective equation of state parameter $w_{eff}$ versus $z$. The red curves correspond to the mimetic $f(R,T)$ in universe filled with collisional matter,
while the blue curves correspond to the mimetic $f(R,T)$ in presence of the non-collisional matter(dust) for the model $\mathcal{Q}(z)= \sqrt{2z+5}$ \cite{baffou29}. }
 \label{fig1}
  \end{figure}

\begin{figure}[h]
\centering
\begin{tabular}{rl}
\includegraphics[width=7cm, height=7cm]{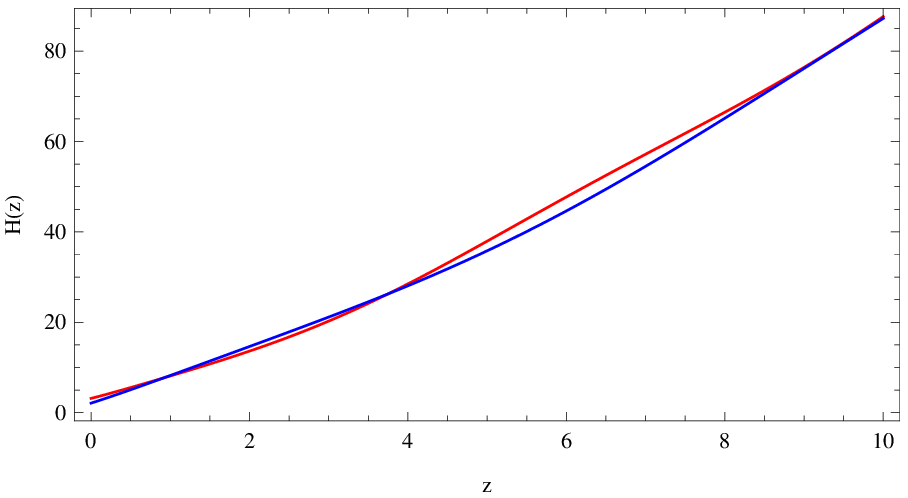}&
\includegraphics[width=7cm, height=7cm]{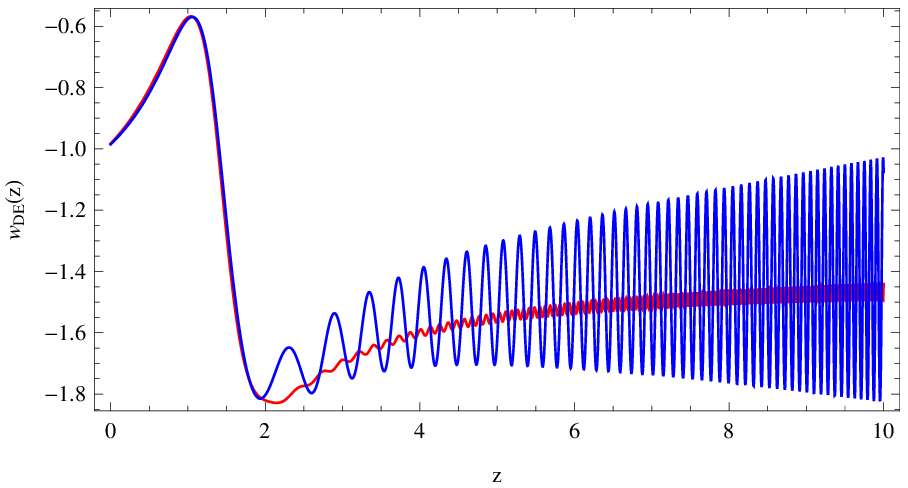}
\end{tabular}
\label{fig2}
\end{figure}
\begin{figure}[h]
\centering
\begin{tabular}{rl}
 \includegraphics[width=7cm, height=7cm]{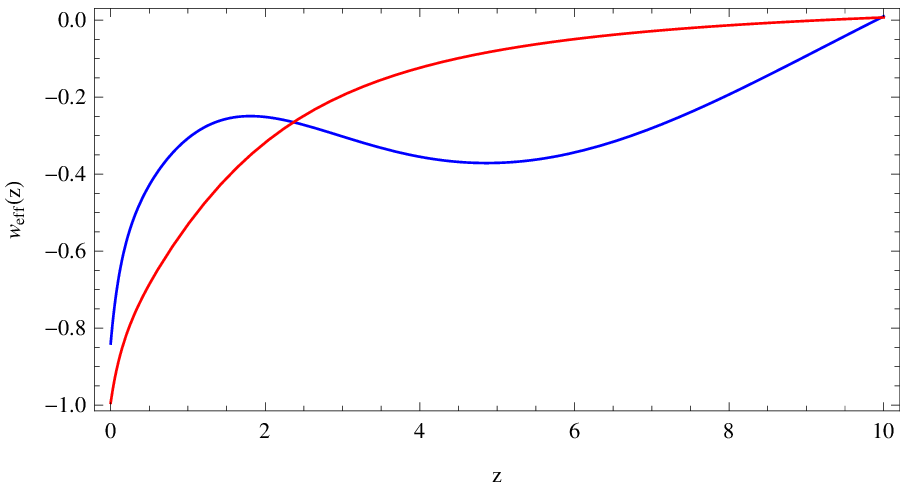}
  \end{tabular}
\caption{Comparison of the Hubble parameter $H(z)$ versus $z$, of the dark energy equation of state parameter $w_{DE}$ versus $z$ and 
of the effective equation of state parameter $w_{eff}$ versus $z$. The red curves correspond to the mimetic $f(R,T)$ in universe filled with collisional matter,
while the blue curves correspond to the mimetic $f(R,T)$ in presence of the non-collisional matter(dust) for the model $\mathcal{Q}(z)= \frac{2z+5}{z+100}$ \cite{baffou29}.
The graphs are plotted for $\Pi_0 = 2.58423$, $\rho_{m0} = 3.1\times10^{-4}$, $\beta = \frac{1+3w}{2(1+w)}$, $\chi = 1.5$, $b= 10^{-5}$  and ${\bar m}^{2}=1.03\times10^{-4}$.}
\label{fig2}
\end{figure}

Concerning the figure (\ref{fig1}), we remark that the Hubble parameter, the dark energy equation of state parameter and the effective equation of state parameter
are practically the same evolution for the each type of matter content considered. We note no significant oscillating behavior at the Hubble parameter and the effective
equation of state parameter level whereas for the dark energy equation of state parameter, it can be seen in mimetic $f(R,T)$ gravity and in presence of the non-collisional matter the oscillations 
occurs when the redshift increase and go toward $z= 3.5 $. For the second figure, we remark that the Hubble parameter and the dark energy equation state parameter
are very similar behavior for the each type of matter content. Moreover at the dark energy equation state level, the oscillating behavior is milder in comparison
to the mimetic $f(R,T)$ gravity in Universe filled with the non-collisional matter case and this oscillations occurs until approximately $z\simeq 2$.
With regards to the effective equation of state parameter, as it can be seen in bottom plot of (\ref{fig2}), the two curves are very similar behavior for the low values
of the redshift but goes toward $w_{eff}=0$ when the redshift $z$  increase.

\section{Conclusion}
In  this paper, we examined the late-time cosmological evolution in the context of mimetic $f(R,T)$ gravity with Lagrange multiplier and mimetic potential. 
Within the assumption that the matter content of the Universe are collisional matter and relativistic matter (radiation), we demonstrated how 
the $f(R,T)$ gravity, the mimetic potential and the Lagrange multiplier affect the late-time cosmological evolution. By solving numerically the main
equation which can be described the cosmological evolution of the dark energy, we focused our attention on the cosmological evolution of the Hubble parameter, 
the dark energy equation parameter and also the effective equation of state parameter in mimetic $f(R,T)$ gravity in Universe filled with collisional matter and we compared 
with the case when the Universe contains essentially the non-collisional matter (dust).
The results obtained showed that in both case and for two $\mathcal{Q}(z)$ models  the curves correspond to the mimetic $f(R,T)$ gravity in presence
with the collisional matter  are in better agreement with the observations data than those obtained in mimetic $f(R,T)$ gravity in presence of the non-collisional matter (dust) and can be used to reduce the amplitude of the dark energy oscillations. Generally the late-time behavior appears and 
the contribution of the collisional matter in mimetic $f(R,T)$ gravity may be considered as alternative method to damp the dark energy oscillations.

 \newpage
 

\begin{thebibliography}{17}
\addcontentsline{toc}{chapter}{Bibliographie}

\bibitem{baffoua} A. G. Riess et al. , Astron. J. 116, 1009 (1998); S. Perlmutter et al., Astrophys. J. 517, 565 (1999); P. de Bernardis et al., Nature 404, 955 (2000).

%


%
%
%
%



\bibitem{baffou5} E. J. Copeland, M. Sami and S. Tsujikawa, Int. J. Mod. Phys. D 15, 1753 (2006); S. Nojiri and S. D. Odintsov, Int. J.
Geom. Meth. Mod. Phys. 4, 115 (2007); T. P. Sotiriou and V. Faraoni, arXiv:0805.1726 [gr-qc].

\bibitem{baffou6} D. A. Felice,  Rev. Rel. 13:3 (2010); P. G. Bergmann, Int. J. Theor. Phys. 1:2536 (1968).




\bibitem{baffou8} S. Tsujikawa, Lect. Notes Phys. 800 (2010) 99; S. Nojiri, S. D.
Odintsov, Int. J. Geom. Meth. Mod.Phys. 4 (2007) 115.



\bibitem{baffou9} M. C. B. Abdalla, S.
Nojiri, Sergei D. Odintsov, Class. Quant. Grav. 22 (2005) L35.

\bibitem{baffou10} A. De Felice, S. Tsujikawa, Living Rev.Rel. 13 (2010) 3; Thomas P. Sotiriou, Class.Quant.Grav.
26 (2009) 152001.

\bibitem{baffou11} S. Nojiri, S. D. Odintsov, Phys.Rev. D70 (2004) 103522;
 E. Elizalde, E.O. Pozdeeva, S. Yu.
Vernov, Phys.Rev. D85, 044002 (2012); G. Cognola, E. Elizalde, Sergio
Zerbini, Phys.Rev. D87 (2013) 4, 044027.

\bibitem{baffou12} K. Bamba,
S. D. Odintsov, D. Saez-Gomez, arXiv:1308.5789, S. Nojiri, S. D.
Odintsov, arXiv:1306.4426.

\bibitem{baffou13} L. Sebastiani, D. Momeni, R. Myrzakulov, S.D. Odintsov,
arXiv:1305.4231; K. Bamba, S.  Nojiri, S. D.
Odintsov, arXiv:1304.6191. 

\bibitem{baffou14} Y. F. Cai, E. N. Saridakis, M. R. Setare and J. Q. Xia, Phys. Rept. 493, 1 (2010),
arXiv:0909.2776.

\bibitem{baffou15} G. Cognola, E. Elizalde, S. Nojiri, S. D. Odintsov, L. Sebastiani, S. Zerbini Phys.Rev.
D77 (2008) 046009.

\bibitem{baffou16} E. Elizalde, S. Nojiri, S. D. Odintsov, L. Sebastiani, S. Zerbini, Phys.Rev. D83 (2011)
086006.

\bibitem{baffou17} T. Harko, F. S. N. Lobo, S. Nojiri and S. D. Odintsov, Phys. Rev.
D 84 (2011) 024020, arXiv: 1104.2669 [gr-qc].




\bibitem{baffou18} M. J. S. Houndjo, Int. J. Mod. Phys. D 21, 1250003 (2012);
M. J. S. Houndjo and O. F. Piattella, Int. J. Mod. Phys. D 21,
1250024 (2012); D. Momeni, M. Jamil, and R. Myrzakulov,
Eur. Phys. J. C 72, 1999 (2012).

\bibitem{baffou19} F. G. Alvarenga, M. J. S. Houndjo, A. V. Monwanou, and
J. B. Chabi-Orou, J. Mod. Phys. 4, 130 (2013).

\bibitem{baffou20} M. Sharif and M. Zubair, J. Cosmol. Astropart. Phys. 03
(2012) 028; 05 (2013) E01; M. Jamil, D. Momeni, and
R. Myrzakulov, Chin. Phys. Lett. 29, 109801 (2012).

\bibitem{baffou21} M. J. S. Houndjo, C. E. M. Batista, J. P. Campos, and O. F.
Piattella, Can. J. Phys. 91, 548 (2013).

\bibitem{baffou22} E. H. Baffou, A. V. Kpadonou, M. E. Rodrigues, M. J. S. Houndjo and J. Tossa,
Space Sci. 355 (2014) 2197.

\bibitem{baffou23} E. H. Baffou, M. J. S. Houndjo, M. E. Rodrigues, A. V. Kpadonou and J. Tossa,
Chinese Journal of Physics 55 (2017) 467–477, 1509.06997 [gr-qc]; E. H. Baffou,  M. J. S. Houndjo and I. G. Salako,
Int. J. Geome Methods Mod. Phys, doi: 10.1142/S0219887817500517.



\bibitem{baffou24} M. E. S. Alves, P. H. R. S. Moraes, J. C. N. de Araujo and M. Malheiro, arXiv: 1604.03874;
C. P. Singh and P. Kumar, arXiv: 1609.01477 [gr-qc].

\bibitem{baffou25} P. H. R. S. Moraes, Eur. Phys. J. C75(4)
(2015) 168, arXiv: 1502.02593 [gr-qc]; H. Shabani  and M. Farhoudi, arXiv: 1407.6187v2. 


\bibitem{baffou26} A. H. Chamseddine and V. Mukhanov, JHEP 1311 (2013) 135
[arXiv:1308.5410 [astro-ph.CO]].

\bibitem{baffou27} S. Nojiri and S. D. Odintsov, arXiv:1408.3561.

\bibitem{baffou28} V. K. Oikonomou, arXiv: 1609.03156v1 [gr-qc].

\bibitem{baffou29} S. D. Odintsov, V. K. Oikonomou, arXiv: 1608.00165v1 [gr-qc].

\bibitem{baffou30} S. Nojiri and Sergei D. Odintsov, arXiv: 1408.3561v3 [hep-th].

\bibitem{baffou31} G. Leona E. N. Saridakisb, arXiv: 1501.00488v3 [gr-qc].

\bibitem{baffou32} A. V. Astashenok, S. D. Odintsov, V. K. Oikonomou,  arXiv: 1504.04861v2 [gr-qc].

\bibitem{baffou33} M. Shiravand, Z. Haghani and S. Shahidi, arXiv:1507.07726v2 [gr-qc].

\bibitem{baffou34} S. D. Odintsov and V. K. Oikonomou, arXiv: 1508.07488v2 [gr-qc].

\bibitem{baffou35} S. Nojiri, S. D. Odintsov and V. K. Oikonomou, arXiv: 1608.07806 v2 [gr-qc].

\bibitem{baffou36} E. H. Baffou, M. J. S. Houndjo,  M. E. Rodrigues, A. V. Kpadonou and J. Tossa, Phys. Review D 92, 
Doi: 10.1103/PhysRevD.92.084043.





\bibitem{a2} A. Golovnev, Phys. Lett. B 728, 39 (2014)
[arXiv:1310.2790].
\bibitem{a3} A. O. Barvinsky, JCAP
1401, no. 01, 014 (2014) [arXiv:1311.3111].
\bibitem{a4} A. H. Chamseddine, V. Mukhanov and A. Vikman, JCAP
1406, 017 (2014) [arXiv:1403.3961].
\bibitem{a5} M. Chaichian, J. Kluson, M. Oksanen and A. Tureanu, [arXiv:1404.4008].
 \bibitem{a6} O. Malaeb, [arXiv:1404.4195].
\bibitem{a7} N. Deruelle and J. Rua, JCAP 1409, 002 (2014), [arXiv:1407.0825].
\bibitem{a8} D. Momeni, A. Altaibayeva and R. Myrzakulov,
[arXiv:1407.5662].
 
\bibitem{b1}S. Nojiri and S. D. Odintsov,
[arXiv:1408.3561].
\bibitem{b2} K. Kleidis and N. K. Spyrou, Astron. Astrophys. 529, A26
(2011).
\bibitem{b3} K. Freese and M. Lewis, Phys. Lett. B 540, 1 (2002).

\bibitem{b4} V. K. Oikonomou and N. Karagiannakis, Classical Quantum
Gravity 32, 085001 (2015).

\bibitem{nojori} S. D. Odintsov, V. K. Oikonomou, arXiv: 1608.00165v1 [gr-qc].
 

\end{thebibliography}
\end{document}